# Oxide-stabilized microstructure of severe plastically deformed CuCo alloys

Martin Stückler[a,1], Jakub Zálešák[a], Timo Müller[b,2], Stefan Wurster[a], Lukas Weissitsch[a], Martin Meier[c,d], Peter Felfer[c], Christoph Gammer[a], Reinhard Pippan[a], Andrea Bachmaier[a3]

[a]*Erich Schmid Institute of Materials Science, Austrian Academy of Sciences, Jahnstraße 12, 8700 Leoben, Austria*
[b]*Deutsches Elektronen Synchrotron, Notkestraße 85, 22607 Hamburg, Germany*
[c]*Department of Materials Science, Institute for General Materials Properties, Friedrich-Alexander University Erlangen-Nürnberg, Martensstraße 5,91058 Erlangen, Germany*
[d]*Department of Materials, University of Oxford, 16 Parks Road, Oxford, OX1 3PH, United Kingdom*

**Abstract**

Nanocrystalline materials are well known for their beneficial mechanical and physical properties. However, it is of utmost importance to stabilize the microstructure at elevated temperatures to broaden the window of application e.g. by pinning grain boundaries through impurities or oxides. CuCo alloys, severe plastically deformed using high-pressure torsion, were used to investigate the evolution of oxides upon annealing. These investigations were performed using electron microscopy, synchrotron high-energy X-ray scattering and atom probe tomography. Monitoring the evolution of oxides by in-situ small angle scattering investigations indicates the growth of primary oxides as well as the formation of another species of oxide at elevated temperatures. Slightly different coarsening can be observed in heat-treated samples, which can be traced back to the influence of different oxide amounts. The initially supersaturated CuCo matrix changes by the decomposition process as well as Co- and Cu-based oxides formation and growth. It has been found that the major

---

[1] Current Address: TDK Electronics GmbH & Co OG, 8530 Deutschlandsberg, Austria
[2] Current Address: Anton Paar GmbH, 8054 Graz, Austria
[3] Corresponding Author: Andrea Bachmaier, andrea.bachmaier@oeaw.ac.at

number of oxides which are present after annealing, can already be found in the starting materials.

*Keywords:* severe plastic deformation, high-pressure torsion, supersaturation, nanocrystalline material, atom probe tomography, small-angle X-ray scattering

## 1. Introduction

Modern materials for energy applications require optimized functional properties while exhibiting enhanced mechanical properties. For a broad variety of applications, nanocrystalline materials are superior with respect to their coarse-grained counterparts. On the one hand grain sizes in the nanometer-regime can give rise to enhanced functional properties [1,2]. Especially for magnetic materials the nanocrystalline regime results in superior properties, e.g. the performance of hard magnetic Nd-Fe-B is increased [3]. Also in nanocrystalline soft magnets a rapid drop in effective magnetocrystalline anisotropy is induced causing a collapse in coercivity and therefore resulting in very soft magnetic materials [4,5]. On the other hand, nanocrystalline materials are known to exhibit enhanced mechanical properties [2,6,7]. Combining the superior functional and mechanical properties, facilitates the synthesis of lighter and more energy-efficient materials [8].

In contrast to many other processing techniques, it has been demonstrated that high-pressure torsion (HPT) [9,10] is capable of producing nanocrystalline bulk specimens. When using powders as starting materials, a broad variety of compositions can be investigated, whereby metastable states can form which even retain after pressure release. Regarding the aspect of functional properties, HPT has been facilitated to process magnetic materials and further tune their properties [11–15].

An important aim in the research of modern magnets is the synthesis of materials for higher operating temperatures [8,16]. Due to the enhanced numbers of diffusion paths and the high number of grain boundaries, nanocrystalline materials usually lack thermal stability. Pure metals, HPT-processed in liquid nitrogen, exhibit grain growth when stored under ambient conditions as has been demonstrated by Edalati et al. [17]. To stabilize nanocrystalline materials, grain growth has to be suppressed which can be facilitated by the addition of second phase particles such as oxides [18,19], which are able to pin grain boundaries [20,21].

In [22], it was shown that a binary powder mixture of Co and Cu exhibits soft magnetic properties which are combined with an extraordinarily high thermal stability of the microstructure. After annealing at 400 °C (ca. 0.4 of homologous temperature $T_{hom,Co}$) the grain size remained stable, even more the material stayed in the ultra-fine grained regime after annealing at 600 °C (ca. 0.5 $T_{hom,Co}$). Additionally, soft magnetic properties prevailed during annealing as the coercivity as a function of annealing temperature was significantly below the value of pure Co for all investigated states.

This study targets at identifying the small particles, stabilizing the microstructure and furthermore to investigate their evolution during annealing treatments. However, these second phase particles are very small and thus challenging to observe. Therefore, high resolution techniques, like transmission electron microscopy (TEM), synchrotron experiments and atom probe tomography (APT) need to be employed.

In this study we used small-angle X-ray scattering (SAXS) to resolve the oxides governing grain growth. This technique is complementary to the microscopy approaches, as it delivers a size distribution of the oxides in a statistically significant volume. On the other hand, the determination of quantitative microstructural information from SAXS data requires additional information and/or assumptions on the sample, in particular for complex nanostructures. The experiment is conducted using a soft magnetic specimen processed by HPT from a powder mixture of Co and Cu. To compare the as-deformed microstructure and to disentangle the effect of initially introduced impurities, a second sample is processed from bulk material. Both samples exhibit a supersaturated microstructure in the as-deformed state. Furthermore, the thermal stability of the powder-based sample is investigated with an in situ annealing treatment using SAXS in combination with wide-angle X-ray scattering (WAXS) to give insights into the evolution of the microstructure during annealing. The results obtained by SAXS are correlated to APT and TEM.

## 2. Experimental

A powder mixture of nominal 70at% Co (Alfa Aesar, -22 mesh, Puratronic, 99.998%) and 30 at% Cu (Alfa Aesar, -170+400 mesh, 99.9%) is consolidated in Ar atmosphere and subsequently deformed by HPT (100 turns at 300 °C and 50 turns at room temperature, RT) resulting in a cylindrical specimen ø8 mm x 0.5 mm and referred to as powder-based sample in the following. Further details regarding sample synthesis are given in [22]. For a second, bulk-based sample an ingot is arc melted five times in Ar-atmosphere (Buehler AM 0,5) using conventional granules (Co: HMW Hauner GmbH & Co. KG 99.99% <15 mm, Cu: HMW Hauner GmbH & Co. KG 99.99% ø3 mm x 3 mm). A cylindrical sample is prepared from the ingot using wire electrical discharge machining and is subsequently deformed by HPT (50 turns at RT) ensuring a saturated microstructural state for radii r ≥ 2 mm. Vickers microhardness was measured with a load of 500 g (HV0.5) using a Buehler Micromet 5104 device. Scanning electron microscopy (SEM) images are obtained using a Zeiss LEO1525 in back scattered electron (BSE) mode.

Synchrotron high-energy SAXS/WAXS experiments are performed at PETRA III beamline P21.2 at Deutsches Elektronen Synchrotron (DESY) [23]. Measurements are performed at 60 keV in transmission mode with the direct beam in axial sample direction at a beam size of 200x200 µm$^2$ (see Ref. [24] for details). An in situ experiment is conducted using an Ar-flushed heating stage (Linkam). The sample is heated at a rate of 10 K/min to a temperature of 500°C. The maximum temperature is kept constant for 1 h, followed by cooling down to room temperature. Temperature calibration is performed by measuring the thermal lattice expansion of a Cu sample under the same conditions. To record the SAXS and WAXS signals simultaneously, three Varex XRD4343CT detectors are used. Geometry calibration for SAXS and WAXS is done using silver behenate and $LaB_6$ respectively. Absolute intensities of the SAXS data are obtained from calibration with a glassy carbon reference sample. SAXS data are analyzed using McSASsoftware [25], calculating a size distribution under the assumption of hard spheres, but without assumptions on the size distribution profile. The McSAS analysis is performed in the radius range 0.9 to 25 nm corresponding to the measured q range. For further analysis of total volume fraction and mean particle sizes, only the radius range 3 to 23 nm is used, ignoring the outer bins of the McSAS size histograms. The outer bins tend to show unrealistic large volume fractions, which can be attributed to background artifacts as well as particles with a size beyond the measured q range.

In a first step, the APT sample was mechanically shaped using a high-accuracy grinding technique [26] followed by electro-chemical refining of the tip and final tip sharpening using a FEI Helios NanoLab 600i focused ion beam (FIB) workstation. APT experiments are performed in laser mode with a pulse repetition rate of 250 kHz, a pulse energy of 75 pJ and a detection rate of 1.5%. The tip volume is reconstructed using the commercial software IVAS version 3.8.2.

For scanning TEM (STEM) investigations, FIB lamellae are machined in a FIB/SEM FEI Helios NanoLab 660. The FIB cutting operations are carried out at accelerating voltages ranging from 30 kV to 2 kV and beam currents ranging from 20 nA to 50 pA. STEM imaging, energy dispersive spectroscopy (EDS) and electron energy loss spectroscopy (EELS) are performed using a FEI Titan Themis 60-300 TEM equipped with ChemiSTEM™ technology and GATAN GIF Quantum energy filter. The TEM was operated at an accelerating voltage of 300 kV and a beam current of 1nA.

## 3. Results and Discussion

*3.1. Structure of powder- vs. bulk-based alloys*

To investigate the influence of starting materials onto the microstructure of HPT-processed samples, two different initial states are used. One sample is processed from a powder mixture, using the same processing route as in [22], the second sample is processed from an arc melted ingot. In that way two different initial states (powder-based and bulk-based) are obtained. In Fig. 1, the Vickers microhardness as a function of increasing radius is displayed. In both samples, the hardness remains nearly unchanged with increasing radii. Thus, a deformation, sufficient to form a microstructurally saturated, steady state has been applied [27]. Fig. 2 (a) and (c) show SEM images of both samples in the as-deformed state. The powder sample features a somewhat smaller grain size than the bulk sample, which is also verified by Vickers microhardness testing. The powder sample exhibits a higher mean hardness (438±7 HV) compared to the bulk-based sample (418±8 HV). Both samples are exposed to annealing treatments (1h, 500 °C) to investigate the influence onto grain growth. Fig. 2 (b) and (d) show the same material after annealing at 500 °C for 1 h. It can be seen that for both samples dark regions start to form, which hints at demixing of phases. Furthermore, tiny spherical particles can be seen in the annealed powder sample and when comparing both annealed samples it is evident from Fig. 2 (b) and (d) that the grain size remains significantly smaller for the powder-based sample. Although powders have been stored and handled in an Ar-filled glove box, residual gases adsorb onto the surface, which is favored by the large specific surface area of the initial powders. These impurities are incorporated into the HPT-processed material and stabilize the microstructure at higher temperatures due to Zener pinning [28]. Furthermore, impurities can affect the saturation grain size reached by HPT-deformation [29,30], which explains the difference in grain size, already in the as-deformed states. For the bulk sample, much larger granules exhibiting a much smaller specific surface area are used. Therefore, less impurities are expected to be in this sample.

Fig. 3 shows synchrotron WAXS patterns of both, the bulk-based and the powder-based specimens in the as-deformed state. The powder-based specimen exhibits a comparatively large peak width indicating again a smaller grain size and/or more crystallographic defects. These results, in combination with SEM images and microhardness measurements from above, indicate a smaller grain size in the powder-based specimen with respect to the bulk-based sample.

Apart from the distinct fcc-pattern, a set of tiny but broad peaks can be identified, which can be assigned to the presence of oxides already in the as-deformed powder-based state. In contrast, the bulk-based sample shows no indications of oxides in the as-deformed state. Due to a similar crystal structure as well as residual stresses being present after sample synthesis by severe plastic deformation (SPD), it is not possible to differentiate between the presence of $Cu_2O$ (space group Pn-3m) or CoO (space group Fm-3m) from the measured WAXS data.

In summary, the results demonstrate a better thermal stability of the powder-based materials compared to bulk-based alloys. This is expected to arise from sample synthesis, in particular from the initial materials, directly causing a larger amount of stabilizing oxides incorporated.

## 3.2. Effect of annealing at atmosphere

To distinguish the effects of impurities already being within the material after HPT-processing from effects of impurities (mostly oxygen) brought into the material during annealing, such treatments are conducted either in a conventional furnace (which is denoted with the prefix "CA" in the following) or in a vacuum furnace (which is denoted with the prefix "VA" in the following). This experiment is exclusively performed on the powder-based specimen. For the annealing treatment CA400 °C, the single phase fcc-structure persists (Fig. 3). A slightly smaller WAXS peak width for the fcc-pattern is observed and a larger signal is obtained in the oxide pattern compared to the as-deformed state. For CA600 °C, two fcc-patterns are present showing demixing of Cu- and Co-rich phases, as has been already described in a previous study [22]. The oxide WAXS peaks show a significant decrease in peak width indicating grain growth. The VA samples do not show large deviations with respect to the CA samples. Since VA samples were exposed to heat during heating, 1h isothermal hold and subsequent cooling, while CA samples were inserted into the pre-heated furnace and taken out after exactly 1 h, the grain growth is somehow more pronounced, i.e. slightly smaller peak widths can be identified for VA samples. However, no apparent deviations in the oxides pattern can be identified between CA and VA samples. This circumstance furthermore demonstrates that in the growth of oxides no ambient oxygen is included, but it is rather controlled by the oxygen available inside the material. Apart from that, it should be mentioned that the hcp-Co phase neither is present in the as-deformed state nor in any annealed state.

As it has been demonstrated that the powder-based sample exhibits the more stable microstructure upon annealing – being of advantage for high temperature applications – the following detailed characterization experiments are performed exclusively on the powder-based specimen. A combination of results from SEM, TEM, APT, SAXS and WAXS was used to investigate the evolution of oxides at elevated temperatures in HPT-processed CuCo materials.

*3.3.*     Atom probe tomography of the as-deformed, powder-based sample

To investigate the elemental distribution in further detail, an in-depth investigation of the powder-based specimen in the as-deformed state is carried out using APT. The resulting mass spectrum indicate the presence of hydrides and oxides, causing several peaks in the mass spectrum to overlap. For the determination of the specimens' composition, peak deconvolution is carried out as described in [31,32]. The analyzed volume is then determined as 71.06 at.% Co, 28.64 at.% Cu and 0.11 at.% O. The remaining 0.19 at.% split up mainly into H and Ga, where the former is a common APT artefact and the latter arises from FIB milling of the APT specimen. Fig. 4(a) shows the reconstructed volume of the as-deformed state. Therein, a fraction of 1% of all Co and Cu atoms is plotted. To investigate the homogeneity, the volume is cut into 100 equally distanced slices in x-, y- and z-direction, whereas the atoms are counted for each slice. In Fig. 4 (b; solid lines) the cumulative Cu count is plotted versus the cumulative Co count, according to the directions of the arrows in Fig. 4(a), following a straight line for each Cartesian axis. This behavior is indicative for a homogeneous distribution of Co and Cu and therefore proves the presence of a supersaturated solid solution. Fig. 4(a) furthermore shows isosurfaces at 5 at.% O. Quite a few small oxides can be identified. The inhomogeneous distribution of O becomes also evident in the ladder diagram in Fig. 4(b; dashed lines). In a further step, the proximity of an oxide is investigated. Fig. 4(c) shows a proxigram of the isosurface, which is highlighted in red in Fig. 4(a). Inside the O-accumulation, the O-composition rises above 20 at.% and, on the other hand, the Cu-composition is found to decrease slightly. We therefore conclude that complex Co- and Cu-containing oxides are present in the as-deformed state.

Due to the nanocrystalline nature of the material, it is conceivable that diffusive processes contribute to the formation of oxides. The diffusion of O from the matrix would cause the formation of an O-depleted zone around the isosurfaces. However, in the proximity histogram (Fig. 4(c)) a constant level of O is found outside the isosurface. Therefore, we assume the oxides did not form due to diffusion at RT but are already present after sample synthesis, i.e. they are either present already in the powder or they are formed during HPT-processing.

*3.4. In situ synchrotron SAXS/WAXS experiments*

To observe the dynamics of the microstructural evolution upon thermal treatment, SAXS and WAXS diffraction patterns are recorded during an in situ annealing treatment.

*3.4.1. Wide angle X-ray scattering*

In Fig. 5, the WAXS patterns are shown as a function of time with the corresponding heating curve. When the highest temperature (500 °C) is reached, still a single-phase fcc-microstructure can be identified. During the isothermal hold, no apparent changes in the fcc-peak pattern can be identified. A slight evolution in height and FWHM of the oxide pattern can be observed. Fig. 6 (b) shows the evolution of the (200)-fcc peak after fitting with a Pseudo-Voigt function. The full width at half maximum (FWHM) shows a decrease upon heating, starting at a temperature of about 100 °C most likely arising from recovery effects at first and slight grain growth. During isothermal hold the FWHM does not change, indicating that no further microstructural changes of the supersaturated phase take place while holding. From the evolution of the peak position during heating, the thermal expansion coefficient for this supersaturated solid solution of CoCu can be obtained by a linear fit to $(13.4 \pm 0.1)\ 10^{-6}\ K^{-1}$ showing a thermal expansion coefficient higher than pure cobalt but lower than pure copper [33,34]. Large residual stresses are present after HPT deformation [35]. During isothermal hold the peak position changes slightly towards larger values. This means that the lattice parameter decreases, indicating a recovery of crystal lattice for example via annihilation of excess vacancies or diffusion of interstitial atoms to grain boundaries. Furthermore, in comparison to the as-deformed state, the lattice parameter is shifted towards a smaller value after thermal treatment which is also indicative for recovery effects.

In Fig. 6(a) the evolution of the oxide peak showing least overlap with the fcc structure is plotted for the in situ treatment. The FWHM starts to decrease rapidly at 100 °C but starts to saturate at about 300 °C. A further but less pronounced decrease in FWHM is recorded during isothermal hold. In the APT analysis it has been demonstrated that Co-based oxides (e.g. CoO) are mainly present in the as-deformed state. It is expected that upon heating the size of these CoO particles increases. The formation of $Cu_2O$ starts at elevated temperatures, as has been reported by various studies [36,37]. Therefore, the further decrease in FWHM might be due to a formation and growth of $Cu_2O$. In summary, the undulating behavior of FWHM in Fig. 6 (a) can be explained by the temperature-dependent formation and growth of at least two types of oxides.

*3.4.2. Small angle X-ray scattering*

In SAXS, the signal arises from differences in electron density on the nanometer scale. The performed SAXS measurements reveal significant scattering intensities for the powder-based sample already in the as-deformed state, increasing significantly during annealing. This intensity is orders of magnitude larger compared to a bulk-based sample. As the electron density in Cu and Co is similar, the contrast cannot be ascribed to Cu nanoparticles in a Co matrix or vice versa. As has been shown above, oxides are already visible in the WAXS data. The SAXS signal is therefore expected to arise from oxides, with the supersaturated CuCo solid solution acting as the matrix. This is also supported by results using McSAS, which lead to volume fractions of Cu well above 100% when assuming the contrast to arise from Cu precipitates embedded in a Co matrix.

Fig. 7 shows the results of SAXS data analysis using McSAS. The calculations are performed with fcc-Co as a matrix and CoO as precipitate, serving as a representative for oxides in general. Fig. 7a shows the calculated radius of the precipitates as a function of time and temperature. The bars represent the standard deviation of the particle size distribution determined by McSAS. The radius increases during heating, starting at a temperature of about 300 °C and continues increasing slowly during isothermal hold. It has to be stated that the histograms on particle size obtained via McSAS indicate that a significant fraction of particles exceeds the restricted size range of this analysis (3 to 23 nm radius) at the higher annealing temperatures. Thus, the actual increase of the mean radius during isothermal hold might be even larger taking into account particles growing beyond the measured size range. The shown data only represents the mean radius of the particles in the analyzed size range.

In Fig. 7b the volume fraction $f_V$ is plotted for the in situ experiment. In this figure, an increase in $f_V$ during heating can be seen. Although a constant or even slightly decreasing $f_V$ during isothermal hold is visible in Fig. 7b, it is expected that the particles grow. As said, particles exhibiting a radius above 23 nm are not resolved and do not contribute to the presented $f_V$.

A comparison of the particle size evolution obtained from SAXS with WAXS data (Fig. 6a) reveals good agreement of both. The discussed enhanced particle growth during isothermal hold, not being represented by the SAXS data due to the limited measured size range, is supported by the steep decrease in FWHM indicating growth of crystallites (see blue curve in Fig. 6(a)). In SAXS, at about 300 °C, a large volume fraction of particles is covered with a concomitant increase in particle size. Also in WAXS, a decrease in FWHM is captured at this temperature and is attributed to the formation of a different oxide species, as mentioned in the section above. This behavior continues throughout thermal hold. Furthermore, the particle sizes obtained from SAXS in as-deformed state are in good agreement with the oxide sizes revealed from APT (see Fig. 4).

Although oxides have been resolved using SAXS, a proper investigation of size and volume fraction necessitates an elaborated experimental setup. The limited size range measured in this experiment limits the quantitative analysis. Extending the measured size range using an experimental setup with better resolution (smaller $q_{min}$ and lower background (e.g. using a hybrid photon counting detector) will facilitate more accurate analysis in the future. Still, it has been demonstrated that SAXS is capable of resolving the evolution of oxides in nanocrystalline materials during in situ annealing treatment, giving valuable complementary information to the WAXS data.

*3.5. Microstructural investigations of the heat treated state*

In the previous section, the evolution of FWHM for the in situ treatment could not be clearly associated to recovery or grain growth. To deliver further insights into the microstructure before and after annealing treatment, these states are investigated using TEM. For not only investigating the oxides but also the evolution of the supersaturated state, a slightly higher annealing temperature of 600°C was chosen – based on results from [22]. Fig. 8(a) shows a high-angle annular dark-field (HAADF) STEM image of the as-deformed powder-based state, revealing the nanocrystalline microstructure. The corresponding EDS maps in Fig.

9(b,c) show the distribution of Co and Cu, respectively. A homogeneous distribution of both elements can be observed, being in good agreement with the APT analysis shown above. Furthermore, stripe-shaped features being characteristic for the synthesis by HPT are visible. In Fig. 8(d), a HAADF-image of the annealed state is shown (CA600 °C, 1h). The microstructure appears coarsened, but nevertheless exhibits grain sizes in the range of about 100 nm. In HAADF, contrast is related to the atomic number. The observed dark particles are therefore most likely oxides. Corresponding EDS maps in Fig. 8(e, f) show the distribution of Co and Cu, respectively. Both images reveal the demixing of phases. In the synchrotron XRD pattern, a single-phase fcc-microstructure has been observed for annealing treatments at 500 °C. In combination with the presented TEM investigations, it becomes evident that the supersaturated solid solution starts to decompose between 500 °C and 600 °C.

The in situ WAXS/SAXS analysis indicated the presence of two different species of oxides. To shed light on the composition of oxides after annealing, EELS investigations are conducted. Fig. 9 and 10 show HAADF-STEM images and corresponding EELS maps for Co, Cu and O for the powder-based sample annealed at 600°C. The majority of oxides is copper-based, but also tiny (most likely) residual Co-based oxides can be found (marked with white circles in Fig. 10(d)). Furthermore, the demixing of the supersaturated solid solution is covered in good detail in these maps. In Fig. 9(b,c) dotted patterns (marked with dashed white circles in Fig. 9(d)) can be observed showing demixed regions. Spinodal decomposition has been observed for various compositions in supersaturated Co-Cu [38,39]. In [22], an increasing coercivity upon annealing has been revealed in supersaturated solid solutions which has been ascribed to the spinodal decomposition. Furthermore, in similarly processed samples an increase in magnetoresistive behavior upon annealing is thought to arise from spinodal decomposition as well [40]. The results obtained by EELS are a strong hint for spinodal decomposition in the investigated Co-Cu supersaturated solid solutions. However, it is rather difficult to provide direct experimental evidence for spinodal decomposition to justify the decomposition mechanism and exclude a classical nucleation and growth mechanism by EELS investigations.

In Fig. 10(c) the O-distribution is measured using EELS. Comparing this data to the Co-distribution, Fig. 10(a), as well as Cu-distribution, Fig. 10(b), it becomes evident that both, Co-rich and Cu-rich oxides are present. These results are in accordance with data presented above, which show a majority of Co-rich oxides in the as-deformed state and a formation of Cu-rich oxides upon annealing.

**4. Conclusion**

In this study, a combination of results from SEM, TEM, APT, SAXS and WAXS was used to investigate the evolution of oxides at elevated temperatures in HPT-processed CuCo materials. It was found that the type of annealing treatment (atmospheric vs. vacuum) does not influence the amount of oxides found in WAXS patterns. Therefore, the effect of atmospheric oxygen during annealing is considered to be negligible. Upon annealing at 500°C existing oxides grow in size but also new oxides start to form. These oxides have been observed using SAXS, whereas this study demonstrates the capability of this method to resolve the evolutions of size and volume fraction of oxides in nanocrystalline materials during in situ annealing treatment. For slightly higher annealing temperature, the Cu-Co matrix changes by decomposition while maintaining a nanocrystalline microstructure at a considerably large homologous temperature. Furthermore, it has been demonstrated that powder-based materials show a better thermal stability compared to bulk-based alloys. This is expected to arise from sample synthesis, being a direct cause of the larger amount of stabilizing oxides incorporated.

**Acknowledgments**

This project has received funding from the European Research Council (ERC) under the European Union's Horizon 2020 research and innovation programme (grant agreement No 757333). We acknowledge DESY (Hamburg, Germany), a member of the Helmholtz Association HGF, for the provision of experimental facilities. Parts of this research were carried out at PETRA III, beamline P21.2. CzechNanoLab project LM2018110 funded by MEYS CR is gratefully acknowledged for the financial support of the measurements/sample fabrication at CEITEC Nano Research Infrastructure.

**Data Availability Statement**

The raw/processed data required to reproduce these findings cannot be shared at this time as the data also forms part of an ongoing study.

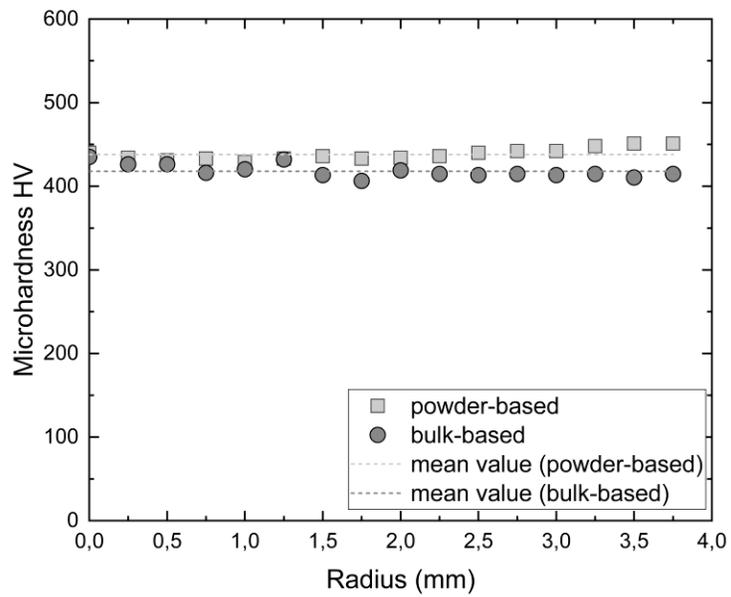

Figure 1: Microhardness of CoCu samples, processed from powders (powder-based) or bulk (bulk-based) as starting materials. A constant microhardness and a microstructurally saturated, steady state has been obtained in both samples.

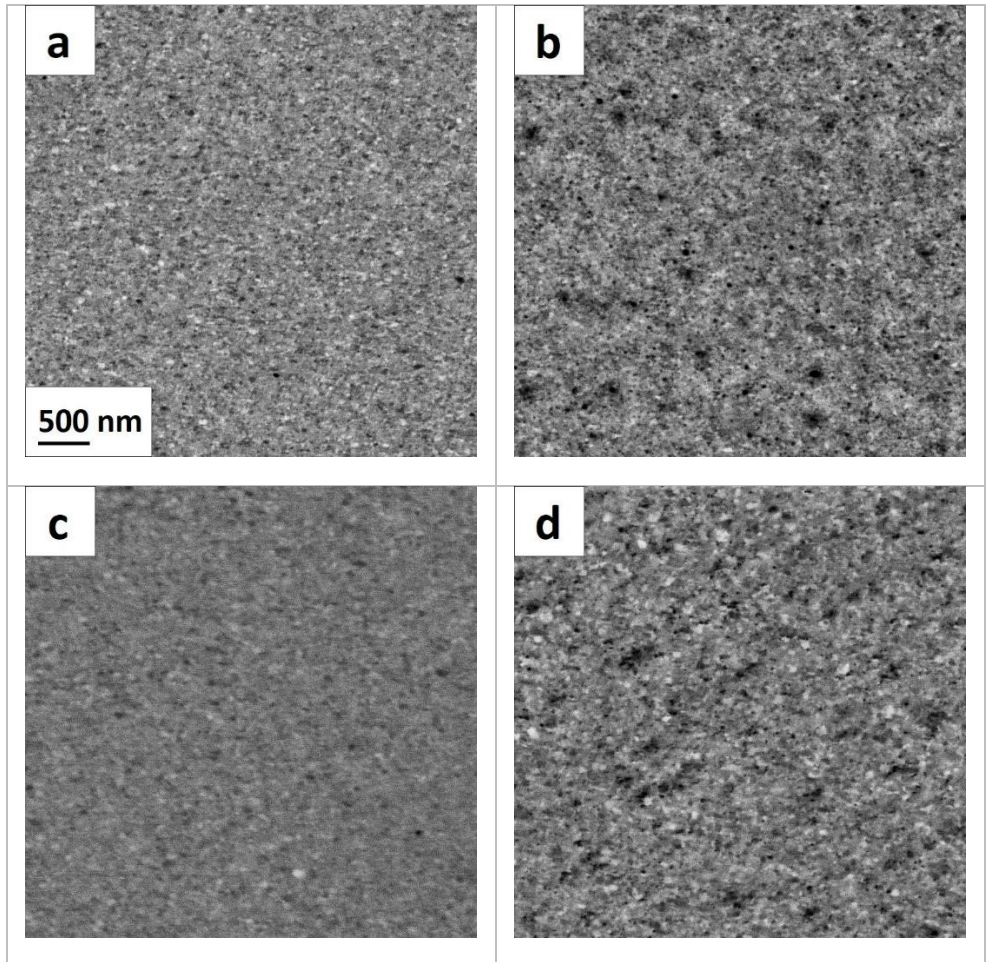

Figure 2: Comparison of SEM images of CoCu samples, processed from powders or bulk as starting materials. (a) and (b) show the powder-based sample in as-deformed and annealed (1h, 500 °C) condition, respectively. (c) and (d) show the bulk-based sample in as-deformed and annealed state (1h, 500 °C), respectively. The micron bar in (a) applies to all images.

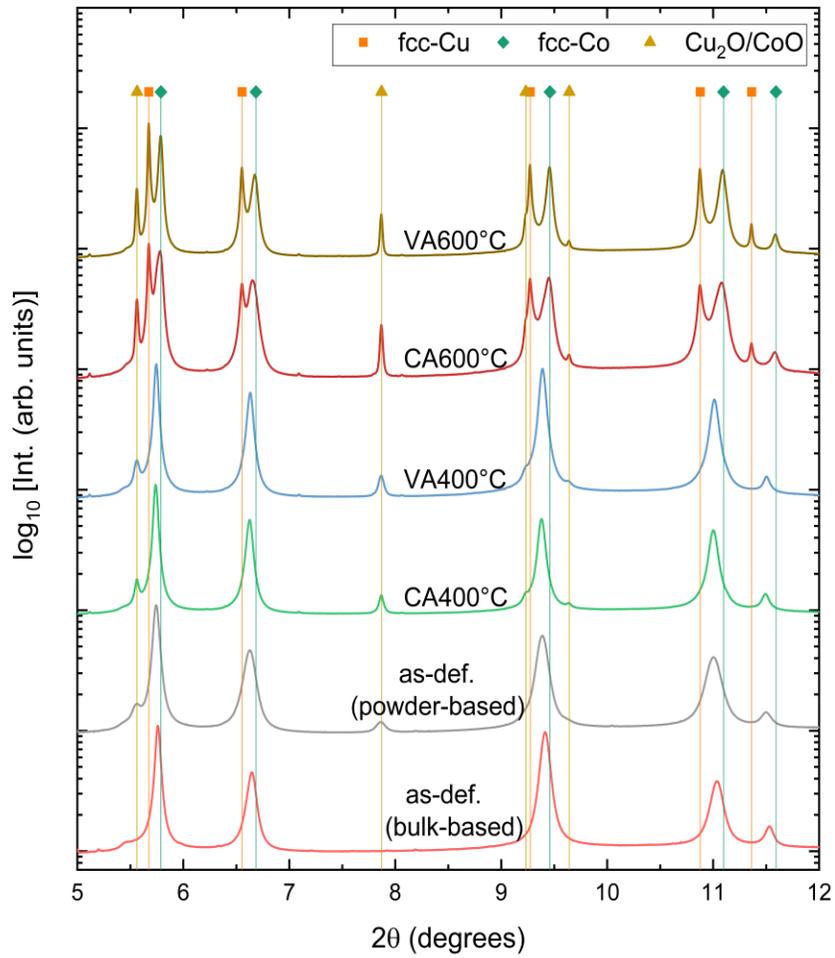

Figure 3: WAXS diffraction patterns of the as-deformed states (bulk-based and powder-based, see Fig. 2) and annealed states (only powder-based). Conventional annealing (CA) and vacuum annealing (VA) treatments were performed at 400 °C and 600 °C for 1 hour. All WAXS measurements are conducted ex-situ at RT.

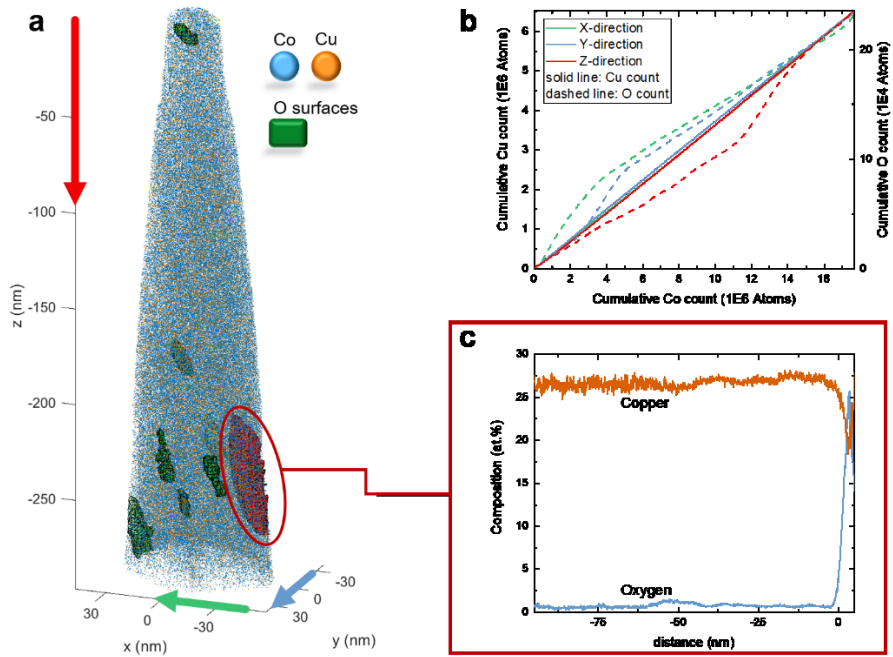

Figure 4: APT reconstruction of the as-deformed state. 1% of all Cu and Co atoms are plotted. Isosurfaces of 5 at.% O are displayed. The reconstructed volume in (a) mainly consists of 71.06 at.% Co, 28.64 at.% Cu and 0.11 at.% O. In (b), ladder diagrams along the colored arrows for Cu and O versus Co are plotted. The proximity histogram in (c) shows the distribution of Cu and O atoms in the proximity of the isosurface, which is highlighted in red in (a).

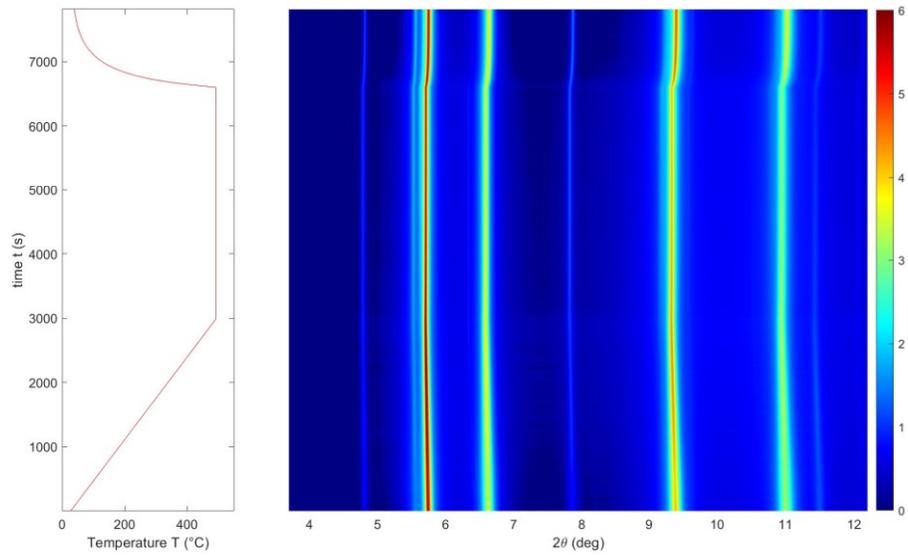

Figure 5: Evolution of WAXS patterns for the in situ experiment. On the left side, the temperature is plotted as a function of time. On the right side, the evolution of the WAXS pattern during the annealing treatment is shown as a color map on square root scale in arbitrary units.

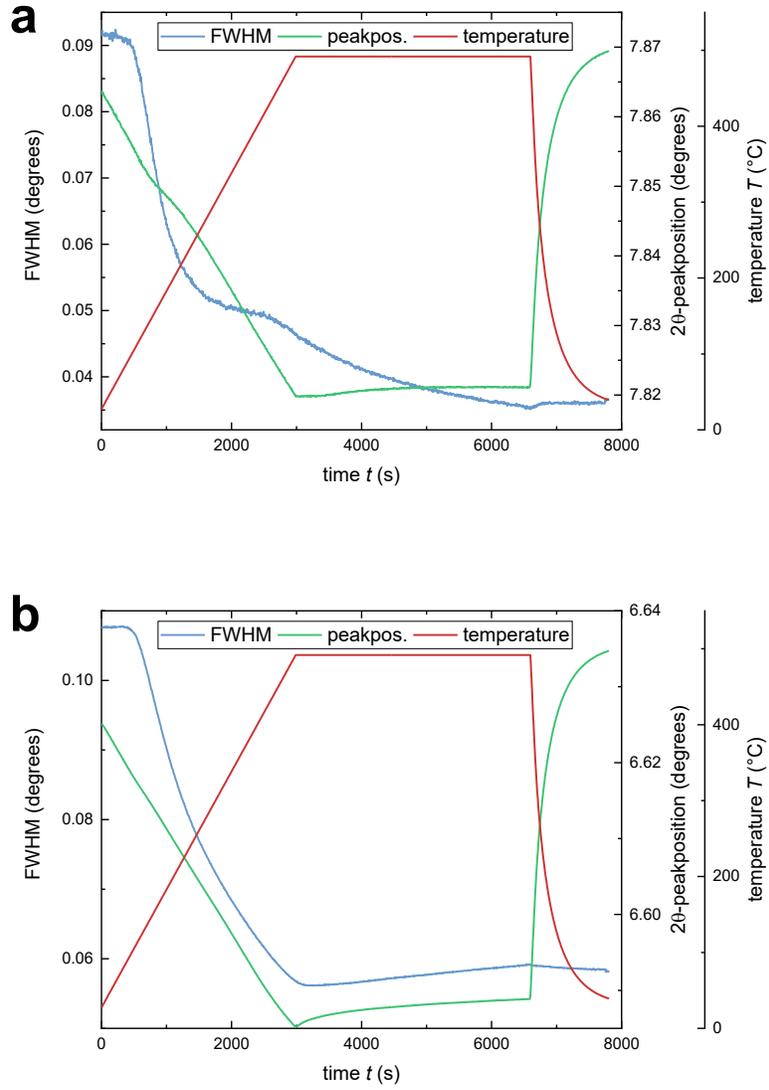

Figure 6: Evolution of full width at half maximum (FWHM), peak position and temperature as a function of time during in situ synchrotron WAXS experiment. In (a) the oxide peak with the least peak overlap with the fcc structure is analyzed. In (b) the (200)-peak of the fcc-supersaturated solid solution is shown.

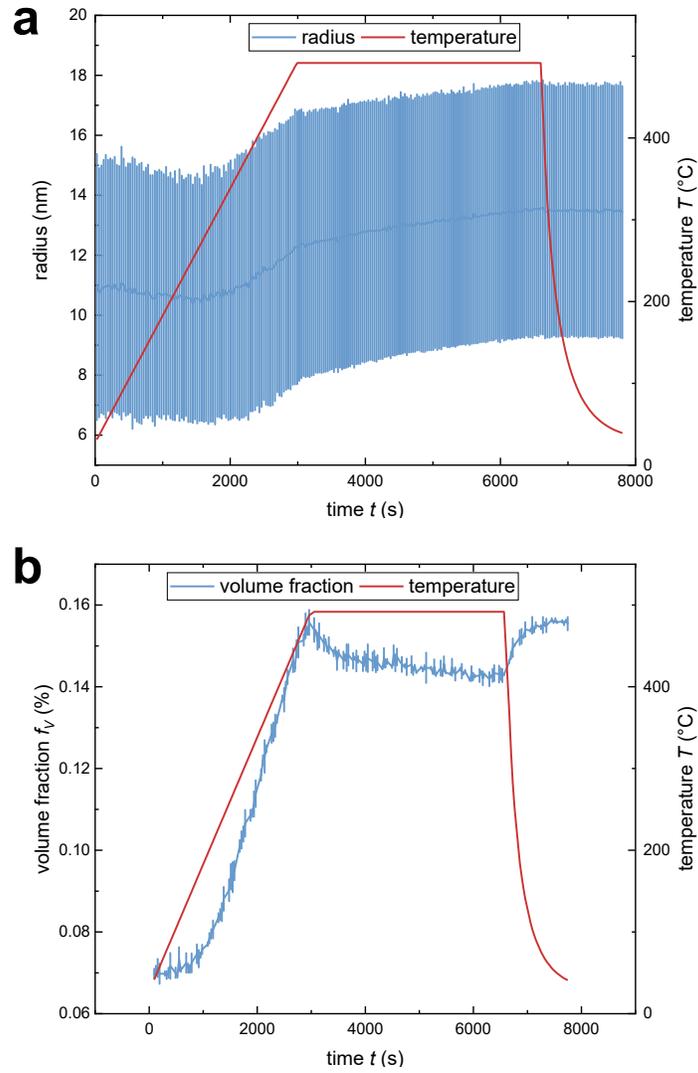

Figure 7: Analysis of in situ SAXS experiments. In (a), the evolution of oxide radius as a function of time is shown. The vertical lines represent the standard deviation of the size distribution. (b) shows the evolution of the oxide volume fraction, including an estimated error, during in situ annealing treatment.

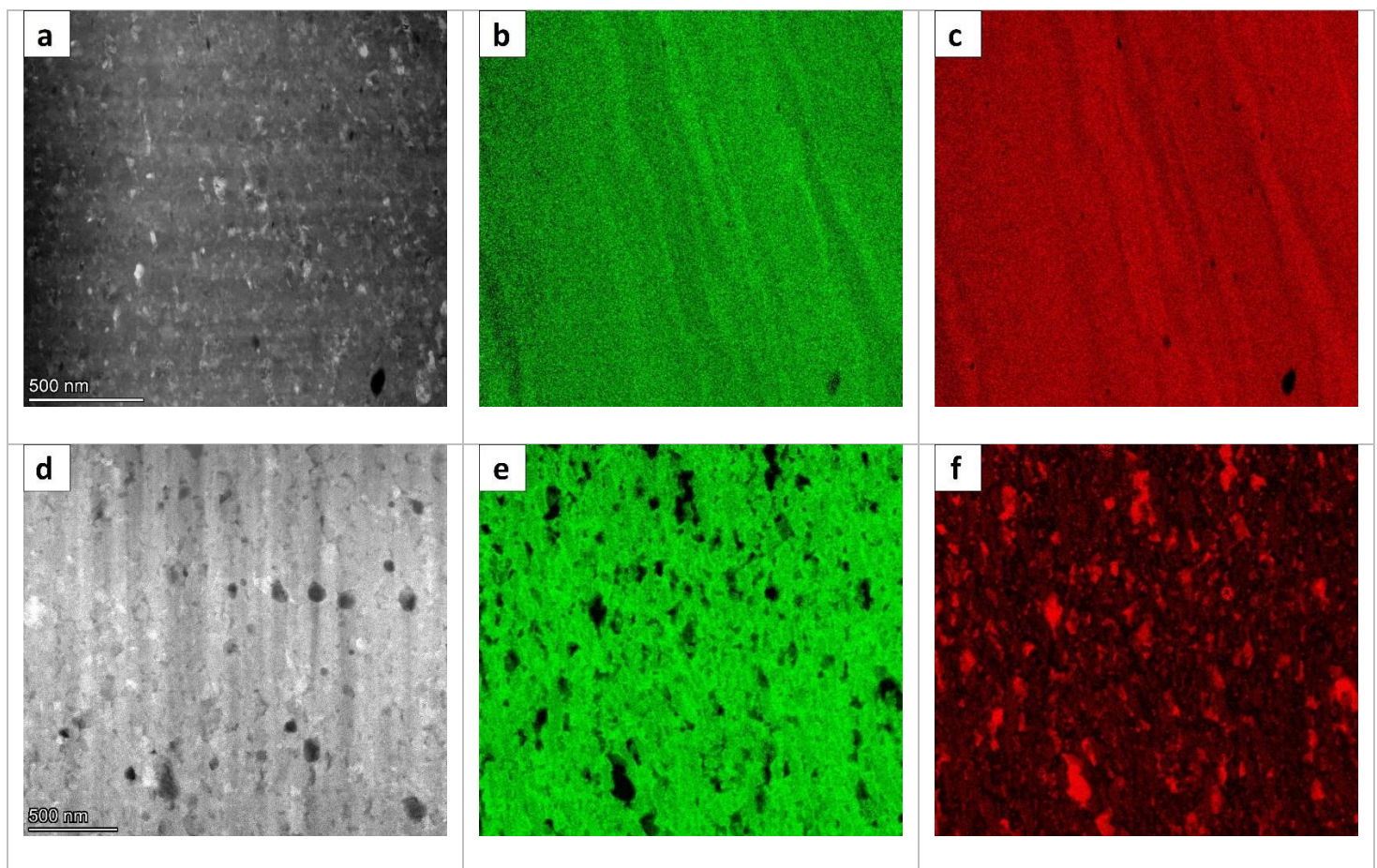

Figure 8: (a) shows a STEM-HAADF image of the as-deformed powder-based sample. In (b) and (c) the corresponding EDS maps show the distribution of Co and Cu, respectively. The scale bar in (a) applies to (b) and (c) as well. (d) shows a STEM-HAADF image for the 600 °C annealed state. (e) and (f) show corresponding EDS maps of Co and Cu, respectively. The scale bar in (d) applies in (e) and (f) as well.

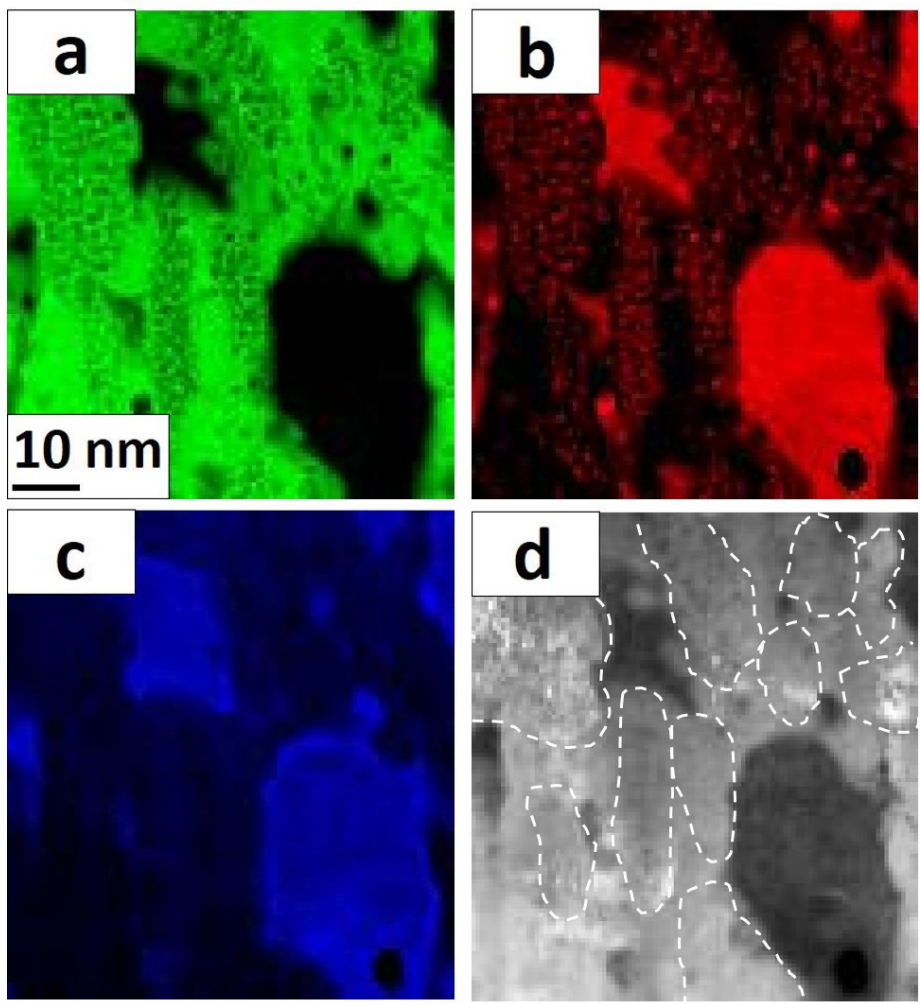

Figure 9: STEM-EELS micrographs of the 600 °C annealed powder-based sample. In (a) and (b) the distributions of Co and Cu, respectively (L-edge for both) are depicted. (c) shows the distribution of O (K-edge). (d) shows a HAADF-STEM image of the area investigated, in which demixed regions are marked with dashed white circles.

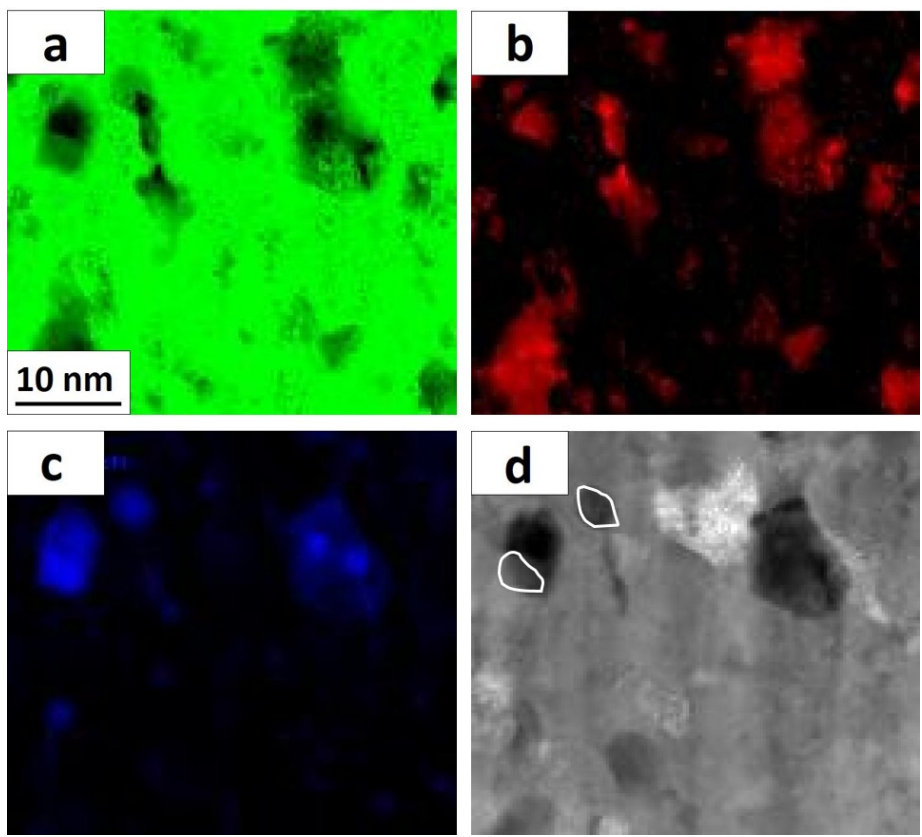

Figure 10: STEM-EELS micrographs of the 600 °C annealed powder-based sample. In (a) and (b) the distributions of Co and Cu, respectively (L-edge for both) are depicted. (c) shows the distribution of O (K-edge). (d) shows a HAADF-STEM image of the area investigated, in which tiny Co-based oxides are marked with white circles.